\documentclass[12pt]{article}

\usepackage[margin=0.9in]{geometry}
\usepackage[]{graphicx}
\usepackage[utf8]{inputenc}
\usepackage{amsmath, titlesec, comment, ccaption, caption, pdfpages, appendix}
\usepackage{optidef}
\usepackage{nameref}

\usepackage[capitalize]{cleveref}
\usepackage[none]{hyphenat}
\usepackage{afterpage}
\usepackage{float}
\usepackage{textgreek}
\usepackage{authblk}
\usepackage{relsize}
\usepackage{algorithm}
\usepackage{algpseudocode}
\usepackage{xurl}
\title{SUND: simulation using nonlinear dynamic models \\  -- a toolbox for simulating multi-level, time-dynamic systems in a modular way}

\author[1]{Henrik Podéus}
\author[2]{Gustav Magnusson}
\author[1]{Sasan Keshmiri}
\author[1,2] {Kajsa Tunedal}
\author[1] {Nicolas Sundqvist}
\author[1,*]{William Lövfors}
\author[1,2,3,*,+]{Gunnar Cedersund}

\affil[1]{Department of Biomedical Engineering, Linköping University, Linköping, Sweden}
\affil[2]{Center for Medical Image Science and Visualization (CMIV), Linköping University, Linköping, Sweden}
\affil[3]{School of Medical Sciences and Inflammatory Response and Infection Susceptibility Centre (iRiSC), Faculty of Medicine and Health, Örebro, Sweden}
\affil[*]{These authors contributed equally to this work}
\affil[+]{Corresponding author}

\date{\today}
\graphicspath{ {./figures/} }

\begin{document}
\pagenumbering{roman}
\maketitle

\clearpage
\begin{abstract}
When modeling complex, hierarchical, and time-dynamic systems, such as biological systems, good computational tools are essential. Current tools, while powerful, often lack comprehensive frameworks for modular model composition, hierarchical system building, and time-dependent input handling, particularly within the Python ecosystem. We present SUND (Simulation Using Nonlinear Dynamic models), a Python toolbox designed to address these challenges. SUND provides a unified framework for defining, combining, and simulating multi-level time-dynamic systems. The toolbox enables users to define models with interconnectable inputs and outputs, facilitating the construction of complex systems from simpler, reusable components. It supports time-dependent functions and piecewise constant inputs, enabling intuitive simulation of various experimental conditions such as multiple dosing schemes. We demonstrate the toolbox's capabilities through simulation of a multi-level human glucose-insulin system model, showcasing its flexibility in handling multiple temporal scales, and levels of biological detail. SUND is open-source, easily extensible, and available at PyPI (\url{https://pypi.org/project/sund/}) and at Gitlab (\url{https://gitlab.liu.se/ISBgroup/projects/sund}).
\end{abstract}

\clearpage
\setcounter{page}{1}
\pagenumbering{arabic}

\section*{Introduction}
\begin{sloppypar}
  Mathematical modeling is an essential tool for understanding complex time-dynamic systems. Such systems are ubiquitous in biology and medicine, where mathematical models are commonly used to study the dynamics of cells, tissues, organs, organisms, and entire populations. Since biological systems are inherently time-dynamic, these mathematical models are typically based on ordinary differential equations (ODEs) or differential algebraic equations (DAEs), enabling simulation of system behavior over time. While mathematical modeling is essential for understanding these systems, their complexity often makes the practical implementation of the simulation and associated model analysis challenging. 

  Many tools are currently available for analyzing ODE and DAE-based models, including frameworks such as SciPy \cite{scipy}, AMICI \cite{amici}, CellML \cite{cellml}, SBToolbox \cite{sbtoolbox}, COPASI \cite{copasi}, CellDesigner \cite{celldesigner}, and Tellurium \cite{tellurium}. These tools provide powerful features for simulating and analyzing dynamic systems within various programming ecosystems. One of the most widely used programming languages within scientific computing and data analysis is Python, as it offers great readability, extensive library ecosystem, and active community support. However, to the best of our knowledge, no single tool within the Python ecosystem simultaneously addresses all of the following requirements: 1) modular model composition with input/output connectivity, 2) creation of multiple model instances within different containers, 3) flexible time-dependent input specification, 4) direct definition of models as ODE/DAE equation systems, and 5) implementation as a free, open-source Python package.

  We therefore present SUND (Simulation Using Nonlinear Dynamic models), an open-source Python toolbox released under the MIT license, designed to fulfill the above listed requirements. SUND provides a unified framework for defining, combining, and simulating multi-level, time-dynamic systems. The toolbox emphasizes modularity: it allows users to define models with inputs and outputs that can be interconnected (\Cref{fig:figure1}A, i)) to build complex systems from simpler components (\Cref{fig:figure1}A, ii)). The interconnected models can have varying temporal scales and differing levels of biological detail, yet can still easily be combined using the toolbox (\Cref{fig:figure1}A, iii)). Furthermore, model instances can be assigned to different containers, allowing for both reuse and isolation of models (\Cref{fig:figure1}A, iv)), useful for applications such as Physiologically Based Pharmacokinetic (PBPK) modeling. Finally, the toolbox supports complex input scenarios mimicking real-world conditions, such as multiple dosing schemes or time-dependent stimulation protocols.
\end{sloppypar}

\section*{Method/Implementation}

SUND is an object-oriented toolbox combining Python and C++ components that leverages the SUNDIALS suite of numerical solvers \cite{sundials} for solving ODE and DAE systems. The user interface is implemented in Python for ease of use, while the core computational functionality is written in C++ to provide efficient computation and direct interface with SUNDIALS solvers. The toolbox uses NumPy \cite{numpy} for numerical operations and Setuptools for C++ compilation. A Python/Setuptools-compatible C++ compiler (GCC, Clang, or MSVC) is required for model compilation.

The toolbox is structured around three main object types: \textit{model} objects, \textit{activity} objects, and \textit{simulation} objects. 

The model object contains the mathematical formulations defined by the user in text files or multiline strings, using syntax similar to SBToolbox \cite{sbtoolbox} to formulate the models. These models are compiled and installed as Python modules that can be used for simulations of various experimental conditions. Users can define model inputs and outputs, making the models modular and suitable for combination with other models or activity objects.

An activity object is an object used to generate inputs to the models. These inputs can be time-dependent functions, including: piecewise constant, piecewise linear, and cubic spline functions, as well as constants. Models can receive user-defined inputs (e.g., external dosing schemes) from these activity objects. The toolbox supports mapping between internal input/output names and external names, facilitating combination of models with different naming conventions.

The simulation object combines model- and activity objects to create a simulation instance, which can be executed over different time intervals with various time steps and numerical solver settings to generate model simulations.

\section*{Results}
\begin{figure}[H]
  \centering
  \makebox[\textwidth]{
    \includegraphics[width=0.9\textwidth]{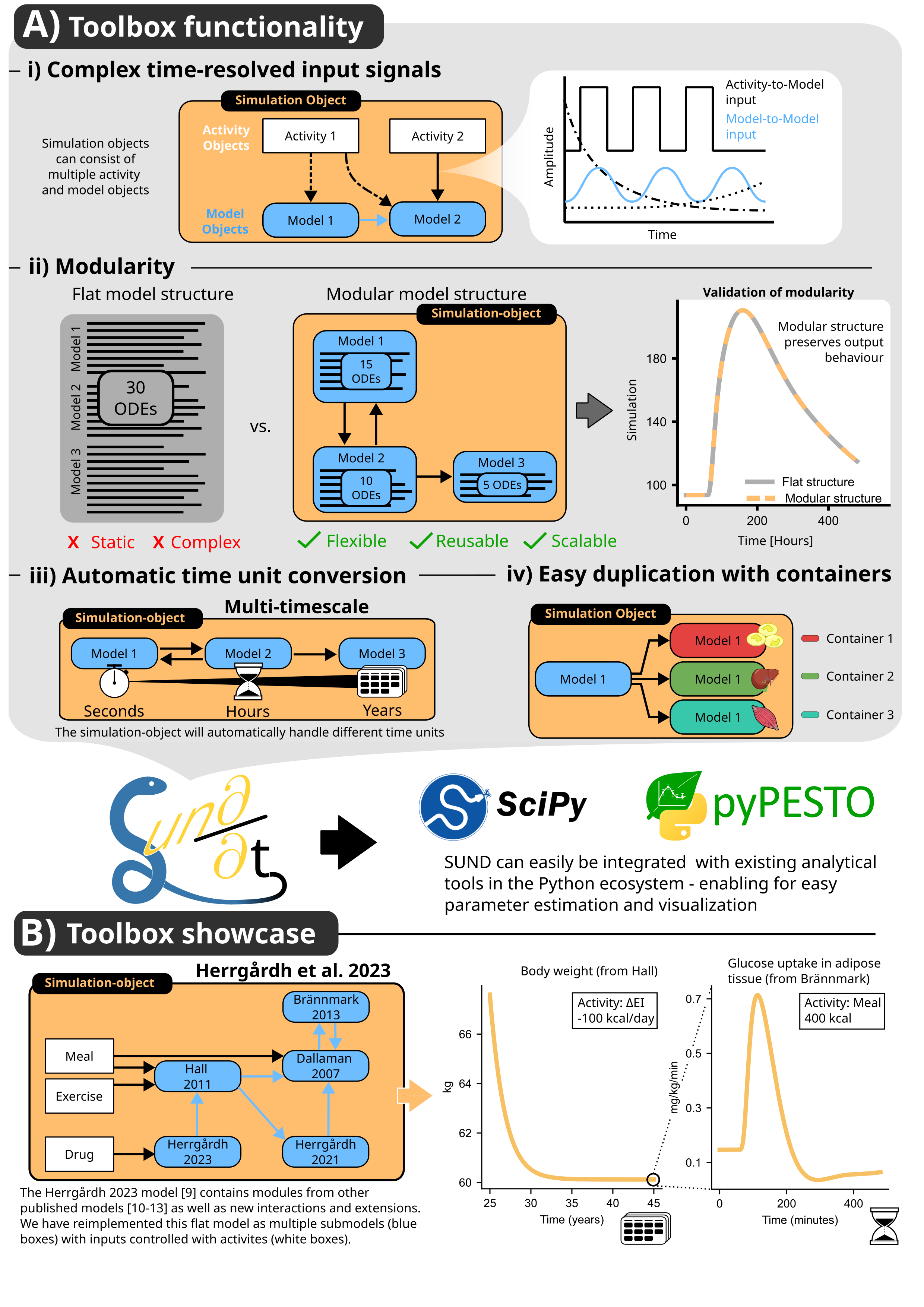}
  }
  \caption{\textbf{Overview of the SUND toolbox capabilities}. The key toolbox features are illustrated in \textbf{A}. \textbf{i}) complex time-resolved input signals can be defined through activity objects, supporting realistic experimental conditions such as dosing protocols and time-dependent perturbations.}
  \label{fig:figure1}
\end{figure}

\begin{sloppypar}
  SUND addresses the challenges of simulating complex multi-level, time-dynamic systems through several key features (\Cref{fig:figure1}). The toolbox's modular design enables users to break down flat models with hundreds of ODEs into manageable, interconnected components (\Cref{fig:figure1}A, ii)). This modularity not only improves maintainability but also facilitates model reuse and collaborative development.

  The input/output connectivity system allows seamless integration of models with different naming conventions through flexible mapping capabilities. Models can receive user-defined inputs from activity objects, for instance: exercise intensity profiles, consumption of nutrients during a meal, or drug administration protocols. This input/output connectivity enables the design of realistic experimental conditions. Default input values can be specified when inputs may not be provided, for example setting a drug concentration to zero when no dosing activity is defined. This design enables robust model composition where components can be easily swapped or modified without breaking the overall system structure. The input/output design provides flexibility for defining complex experimental protocols and realistic simulation of dosing regimens, meal patterns, or other time-varying perturbations.

  For non-smooth inputs (e.g., piecewise constant), simulation objects automatically reinitialize numerical integration at discontinuity points to prevent integrator problems. This relieves users from tracking discontinuities and manually reinitializing integration by subdividing simulations.
\end{sloppypar}

The automatic time unit conversion feature allows seamless combination of models operating across vastly different temporal scales - from nanoseconds to years (\Cref{fig:figure1}A, iii)). Model-, activity-, and simulation objects each specify their individual time units, and during simulation, components are automatically converted to the simulation object's time unit. This capability is particularly valuable in biological modeling where cellular processes may occur on timescales of seconds to minutes, while physiological adaptations happen over days to years, eliminating a common source of errors.

The possibility to assign model instances to different containers simplifies the reuse of sub-models in a larger context, and facilitates isolation of the container-assigned models (\Cref{fig:figure1}A, iv)). Container assigned models can still easily get inputs from other containers as needed by specifying from which container to get the input from.

We demonstrate these capabilities through comprehensive examples available in the online documentation at \url{https://isbgroup.eu/sund-toolbox/}, including a detailed multi-level human glucose-insulin system model based on Herrgårdh et al. \cite{herrgardh_2023}, which itself incorporates components from Dalla Man et al. \cite{dallaman_2007}, Hall et al. \cite{hall_2011}, Brännmark et al. \cite{brannmark_2013}, and Herrgårdh 2021 et al. \cite{herrgardh_2021}. In our example, we demonstrate how this originally flat model structure can be decomposed into interconnected modular components using SUND's framework (\Cref{fig:figure1}B). Additional code examples and executable demonstrations showcase the toolbox's capability to handle models from different sources with varying complexity and timescales.  

\noindent
\textbf{ii}) the toolbox supports modular model composition where complex systems can be built from simpler components, avoiding flat structures with hundreds of ODEs. \textbf{iii}) automatic time unit conversion enables seamless integration of models operating at different timescales (nanoseconds to years). \textbf{iv}) easy assignment to containers allows straightforward creation of multi-container models. \textbf{B} presents a showcase of the toolbox's capabilities, where a complex multi-level human glucose-insulin system model is constructed from interconnected modular components. The model integrates processes across different biological levels, from cellular signaling to whole-body glucose regulation, which is presented by the two plotted features with different time scales originating from different submodules.

\section*{Discussion}
\begin{sloppypar}
  We have presented a toolbox for simulating complex time-dynamic systems, particularly biological systems, using a modular approach with flexible time-resolved inputs. The toolbox enables users to specify ODE and DAE-based models in an accessible format and simulate these models under diverse realistic conditions.

  While designed for user-friendliness and flexibility, SUND has limitations. The toolbox does not provide internal algorithms for parameter estimation or Hessian approximation; however, it integrates easily with external tools such as SciPy's optimization modules \cite{scipy}. Additionally, while SUND handles models and activities across different timescales and supports automatic time unit conversion, it does not provide automatic conversion between non-temporal units (e.g., concentration to molecular amount).

  The modular design philosophy of SUND addresses a significant gap in current Python-based simulation tools by enabling intuitive construction of complex systems from reusable components. This approach is particularly valuable for multi-scale biological modeling, where different subsystems may be developed independently and later integrated. The activity-based input system provides flexibility for defining complex experimental protocols, making SUND well-suited for both research and educational applications.
  
\end{sloppypar}

Future development directions include expanding unit conversion capabilities, integrating methods to solve the sensitivity equations to estimate the Hessian, and developing graphical user interfaces for model construction and visualization.

\section*{Code availability}
The package is available for installation from PyPI (\url{https://pypi.org/project/sund/}), and the source code is available at \url{https://gitlab.liu.se/ISBgroup/projects/sund}. The package is licensed under the MIT license.

\section*{Funding information}
GC acknowledges support from the Swedish Research Council (2023-03186, 2023-05460), the Horizon Europe project STRATIF-AI (101080875), VINNOVA (VisualSweden), ALF (RÖ-1001928), and the Exploring Inflammation in Health and Disease (X-HiDE) Consortium - a strategic research profile at Örebro University funded by the Knowledge Foundation (20200017).

The funders had no role in study design, data collection and analysis, decision to publish, or preparation of the manuscript.

\section*{Author contributions}
HP wrote and revised the manuscript, wrote the code and ran the analysis. 
SK wrote the code, and conceptualized and designed the work. 
GM revised the manuscript, wrote the code, and conceptualized and designed the work. 
KT revised the manuscript, and contributed with examples to the documentation
NS revised the manuscript, and contributed with examples to the documentation
GC revised the manuscript, supervised GM, conceptualized and designed the work, and provided funding. 
WL wrote and revised the manuscript, supervised SK and HP, wrote the code and ran the analysis, and conceptualized and designed the work. 
All authors have approved the final version of the manuscript. 

\section*{Competing interests}
GC is the owner of the company \textit{Aktiebolaget SUND sound medical decision} which have been using the toolbox internally during the development of the toolbox. The rest of the authors declare no competing financial or non-financial interests.

\clearpage
\bibliography{references}

\end{document}